\documentclass[12pt]{iopart}
\input psfig
\usepackage{iopams}

\newcommand{\be}{\begin{eqnarray}}
\newcommand{\ee}{\end{eqnarray}}
\newcommand{\bE}{{\bf E}}
\newcommand{\bH}{{\bf H}}
\newcommand{\bA}{{\bf A}}
\newcommand{\bD}{{\bf D}}
\newcommand{\bj}{{\bf j}}
\renewcommand{\br}{{\bf r}}
\newcommand{\bR}{{\bf R}}
\newcommand{\bk}{{\bf k}}

\newcommand{\bv}{{\bf v}}
\newcommand{\bl}{{\bf l}}
\newcommand{\rot}{{\rm rot}}
\newcommand{\im}{{\rm Im}}
\newcommand{\re}{{\rm Re}}
\newcommand{\cD}{{\cal D}}
\newcommand{\cH}{{\cal H}}
\newcommand{\cG}{{\cal G}}

\newcommand{\zb}{\overline{\zeta}}
\newcommand{\db}{\overline{\delta}}

\begin{document}

\title[Decoherence of electron beams]
{Decoherence of electron beams by electromagnetic
field fluctuations}
\author{Yehoshua Levinson}
\address {Department of
Condensed Matter Physics,\\ The Weizmann Institute of Science,
Rehovot 76100, Israel}

\begin{abstract}
Electromagnetic field fluctuations are responsible for the destruction
of electron coherence (dephasing) in solids
and in vacuum electron beam interference. The vacuum fluctuations are
modified by conductors and dielectrics, as in the Casimir effect,
and hence, bodies in the vicinity of the beams can influence
 the beam coherence. We calculate the quenching of
interference of two beams moving in vacuum parallel to a
thick plate with permittivity
$\epsilon(\omega)=\epsilon_{0}+i 4\pi\sigma/\omega$.
In case of an ideal conductor or dielectric $(|\epsilon|=\infty)$
the dephasing is suppressed when the beams are
close to the surface of the plate, because the random
 tangential electric field $E_{t}$, responsible for dephasing,
is zero at the surface.
The situation is changed dramatically when
 $\epsilon_{0}$ or $\sigma$ are finite. In this case there exists
a layer near the surface, where the fluctuations of $E_{t}$ are
strong due to evanescent near fields.
The thickness of this near - field layer is of the order of the wavelength
in the dielectric or the skin depth in the conductor, corresponding to
a frequency which is the inverse electron time of flight from the
emitter to the detector. When the beams are within this
layer their dephasing is enhanced and for slow enough electrons can be
even stronger than far from the surface.
\end{abstract}


\submitto{\JPA}


\section{Introduction}
\label{sec:Intro}
Quantum electromagnetic (EM) field fluctuations are well known as being
responsible for the Casimir forces, see for example \cite{BMM01}. Less
known is the role of these fluctuations in destructing electron
coherence. In weak localization phenomena
in solids EM fluctuations are one
of the dephasing mechanisms of conduction electrons \cite{AAK82}, see also
\cite{SAI90}. The interference of vacuum electron beams,
observed experimentally \cite{HKS00,T00},
is also quenched by EM fluctuations \cite{F93}, see also \cite{BP01}.
These two papers consider EM fluctuations in vacuum or when  ideal
conductors are present in the vicinity of the beams.
The role of dissipation was discussed in Ref. \cite{APZ97}.
Based on physical arguments, the decoherence
was related to the deceleration of an electron from the beam
due to the Ohmic dissipation of the current produced
in the metal by the image charge.

The aim of this paper is to extend  the calculations
of Refs. \cite{F93,BP01} to the case where the beams
are close to {\it dissipative} bodies and to consider
in detail the experiment geometry when
two interfering beams move in vacuum parallel to a thick  infinite
plate with  permittivity
$\epsilon(\omega)=\epsilon_{0}+i\;4\pi\sigma/\omega$.
Calculations of the dephasing factor in this geometry demonstrate
the crucial role of dissipation.
If the plate is an ideal conductor, $\sigma=\infty$, the fluctuations
of the tangential electric field  $E_{t}$, which are responsible for beam
dephasing in this geometry, are suppressed near the plate surface
because of the boundary condition $E_{t}=0$ at the surface.
However, when $\sigma$ is finite, very strong
fluctuations of $E_{t}$ exist near the plate surface, within
a layer of the order of the skin depth. These near-field fluctuations
dramatically enhance the beam dephasing.
Unexpectedly, a similar effect exists also near
a lossless dielectric with high permittivity, $\sigma=0,\epsilon_{0}\gg 1$,
within a layer of the order of the wave length in the dielectric.

The paper is organized as follows. In Sec.\ref{deph} we
 present the dephasing factor $e^{-K}$ in terms
of the EM field correlator $\cD$ in the case of no dissipation,
Eq.(\ref{Kgen}), and give reasons
why the quantum Langevin equation for the EM field
has to be used when dissipation is present.
In Sec.\ref{Langeq} we derive the Langevin equation  and prove
that the expression of $K$ in terms of $\cD$ is valid in the case of
dissipation too. In this section we present also the relation between
$\cD$ and the EM field retarded Green function $\cG$, which is used to
calculate $\cD$.
The above mentioned special geometry is considered in
Sec.\ref{thickpl}, where $K$ is given as an integral, Eq.(\ref{Kp}),
over wave vectors and frequencies, containing
the spectral density of the EM field fluctuations
$\langle E_{t}E_{t}\rangle_{\bk\omega}$,
and the spectral density $|(\bl)_{\bk\omega}|^2$
of the EM field radiated by the beam electrons.
$\langle E_{t}E_{t}\rangle_{\bk\omega}$
is calculated in Sec.\ref{SpDensg} and Sec.\ref{SpDensS}, where
Eqs.(\ref{Sdiel}) and (\ref{Scond})
demonstrate the enhancement of fluctuations due to near fields.
In Sec.\ref{Deph/nf} we present a model for $|(\bl)_{\bk\omega}|^2$
and calculate explicitly
$K$ as a function of the distance of the beams from the plate $d$
 and the  electron velocity $v$
(see  Eqs.(\ref{A}),(\ref{B}) and (\ref{C}) and the text which follows).
It turns out that the dephasing
enhancement due to near fields is appreciable when  $v$
and $\sigma$ are not very large.
In Sec.\ref{diss-deph}
we discuss the relation between beam dephasing and beam EM radiation.
The Appendix contains some calculation details.
\section{Beam dephasing}
\label{deph}

If one ignores the interaction of the beam electrons with the
EM field, the number
of electrons measured in the interference experiment is
$
n=|\psi_{1}|^2+|\psi_{2}|^2+2{\rm Re}(\psi_{1}\psi_{2}^*),
$
where $\psi_{1}$ and $\psi_{2}$ are the wave-functions corresponding to the
coherent motion of the
electrons in beams 1 and 2, and $n$ is calculated at the detector position.
The interaction with the EM field does not affect the
squares $|\psi_{1}|^2$ and $|\psi_{2}|^2$ (since it does not change the
number of electrons in the beams), but
the product $\psi_{1}\psi_{2}^*$, responsible for the interference pattern,
is multiplied by a factor $e^{i\phi}e^{-K}$ with real $\phi$ and positive $K$.
The first factor only shifts the interference pattern in space, while the
second one reduces the amplitude of the interference oscillations
(compared to the background $|\psi_{1}|^2+|\psi_{2}|^2$) and describes
 dephasing.

To calculate  the strength of dephasing we use the "trace of the environment"
picture \cite{SAI90}.
At $t=t_{0}$, when the electron is emitted from the source, the environment is
in state $|t_{0}\rangle$.
While moving, the electron interacts with the
environment
and perturbs its state. When the electron moving in beam 1 arrives the detector
at time $t_{1}$, the environment evolves due to this interaction
to state $|t_{1}\rangle$. In a similar way one
defines  the state $|t_{2}\rangle$. According to the "trace of the environment"
picture
$e^{i\phi}e^{-K}=\langle t_{2}|t_{1}\rangle.$

One can present the final states of the environment in terms of
evolution operators,
\be
|t_{1}\rangle =U_{1}|t_{0}\rangle,
\qquad U_{1}={\cal T}\exp\left[-\frac{i}{\hbar}\int_{t_{0}}^{t_{1}}
dt H_{1}(t)\right],
\ee
where ${\cal T}$ means time ordering and $H_{1}(t)$ is
the interaction of the electron in beam 1 with the environment.
$H_{1}(t)$ is in the interaction representation,
i.e. sandwiched with evolution exponents $\exp[-i\cH t/\hbar]$
containing the beam electron
Hamiltonian and the environment Hamiltonian.
In a similar way one defines $U_{2}$
in terms of $H_{2}(t)$ and finds
$
\label{12}
\langle t_{2}|t_{1}\rangle =\langle t_{0}|U_{2}^{-1}U_{1} |t_{0}\rangle.
$
For an EM environment, choosing a gauge with zero scalar potential, we have
\be
\label{int}
H_{1}(t)=-\frac{1}{c}\int d\br\,\bj_{1}(\br,t)\bA(\br,t),
\ee
where $\bj_{1}$ electron current density operator for the electron in
beam 1
(sandwiched with the evolution exponents containing the
 beam electron Hamiltonian) and $\bA$ the vector potential operator
 (sandwiched with the evolution exponents containing the EM
 environment Hamiltonian). $H_{2}(t)$ is defined similarly with the
 current $\bj_{2}$.

 In this approach one assumes that at the initial moment $t_{0}$ the
 electron source and the environment are un-correlated. It is also
 assumed that the renormalization of the bare electron parameters
due to the interaction with the EM environment \cite{BP01}
 does not influence
 substantially the dephasing phenomena.

To proceed we  assume,
following \cite{F93},  that the current is a classical quantity.
When there is no dissipation in the EM environment, its Hamiltonian is simply
the EM field Hamiltonian and
 the EM field can be quantized expanding it in normal modes.
It is well known that in this case
 the commutator
$[\bA(\br,t),\bA(\br',t')]$ is an {\it imaginary c-number}, and due
 to the classical nature of the currents $\bj_{1}$ and $\bj_{2}$
 the commutators of $H_{1}(t)$ and $H_{2}(t')$ have the same property.
 Because of this property
the time ordering affects only the phase of the evolution operators \cite{G63},
and one can obtain
\be
U_{1}=e^{i\phi_{1}}V_{1}, \qquad
V_{1}=\exp\left[-\frac{i}{\hbar}\int_{-\infty}^{\infty} dt H_{1}(t)\right],
\ee
if $H_{1}$ is defined to be zero for $t<t_{0}$ and for $t>t_{1}$.
The phase $\phi_{1}$ contains the
commutator $[H_{1}(t),H_{1}(t')]$.
 Defining $H_{2}(t)$ in a similar way, we have
\be
\hspace{-1.7cm}
U_{2}^{-1}U_{1}=e^{i(\phi_{1}-\phi_{2})}V_{2}^{-1}V_{1}
=e^{i(\phi_{1}-\phi_{2})}e^{i\chi}
\exp\left[-\frac{i}{\hbar}\int_{-\infty}^{\infty} dt (H_{1}(t)-H_{2}(t))\right],
\ee
where the additional phase $\chi$ contains the commutator $[H_{1}(t),H_{2}(t')]$.
Averaging this over $|t_{0}\rangle$ we find
\be
\hspace{-1.5cm}
\label{tt}
\langle t_{2}|t_{1}\rangle &=& e^{i\phi}e^{-K}
=e^{i(\phi_{1}-\phi_{2}+\chi)}
\left\langle \exp\left[\frac{i}{\hbar c}\int _{-\infty}^{\infty} dt
\int d\br\; \bj_{12}(\br,t)\bA(\br,t)\right]\right\rangle,
\ee
where $\bj_{12}(\br,t)=\bj_{1}(\br,t)-\bj_{2}(\br,t)$ and $\langle ... \rangle$
means average over $|t_{0}\rangle$. When the initial state of the environment
is an equilibrium state with temperature $T$, the average
 means a thermal average $\langle ...\rangle_{T}$.

The second important property of $\bA (\br,t)$ in the case of no dissipation
is that it is a Gaussian operator with respect to thermal averaging
$\langle ...\rangle_{T}$. After expanding $\bA$ in normal modes this
property follows from the relation \cite{G63}
\be
\hspace{-1.7cm}
\langle\exp(\alpha^{*} a^{\dagger}-\alpha a)\,\rangle_{T}&=&
\exp\left[\frac{1}{2}\langle(\alpha^{*}
a^{\dagger}-\alpha a)^{2}\rangle_{T}\right]
=\exp\left[-|\alpha|\,^2\left(n+\frac{1}{2}\right)\right],
\ee
where $a^{\dagger}$ is the bosonic operator creating a photon in some
normal mode, $n=\langle a^{\dagger}a\rangle_{T}$ is the occupation
number of this mode, and $\alpha$ is a complex number.
Using the Gaussian properties of $\bA$
 one can perform the thermal averaging in Eq.(\ref{tt})
and obtain
\be
\label{Kgen}
\hspace{-1.5cm}
K=\frac{1}{2(\hbar c)^2}\int dt dt'\int d\br d\br'
j_{12}^{\alpha}(\br,t) j_{12}^{\beta}(\br',t')
\langle A_{\alpha}(\br,t)A_{\beta}(\br',t')\rangle_{T},
\ee
where $\alpha,\beta=x,y,z$.
If one defines the thermal correlator
\be
\label{DT}
\cD _{\alpha\beta}(\br,\br'; t-t')
=\frac{1}{2}
\langle A_{\alpha}(\br,t)A_{\beta}(\br',t')+
A_{\beta}(\br',t')A_{\alpha}(\br,t)\rangle_{T},
\ee
 the final result is
\be
\label{Ktemp}
K=\frac{1}{2(\hbar c)^2}\int dt dt'\int d\br d\br'
j_{12}^{\alpha}(\br,t) j_{12}^{\beta}(\br',t')
\cD _{\alpha\beta}(\br,\br'; t-t').
\ee
It was obtained for $T=0$ in Ref.\cite{F93} and for $T\neq 0$
in Ref.\cite{BP01}. We derived it in a different way to emphasize
the two assumptions under which this result
is valid (for classical currents), namely:
(i) the commutator of the field operator $\bA(\br,t)$ is
an {\it imaginary $c$-number}
and (ii) $\bA(\br,t)$ is a {\it Gaussian quantity with respect to thermal
averaging}.

When dissipation is present,
the EM environment Hamiltonian includes not only the EM field,
but also the electrons in the absorbing bodies and their interaction
with the EM field. If the field operator $\bA$
is defined as sandwiched by evolution exponents
containing  the EM field Hamiltonian {\it only}, it has to be considered as a
{\it random}
quantity due to the influence of the dissipative electron system
in the absorbing bodies. These electrons are the thermal bath, whose
temperature defines the temperature of the EM field.
Being a random operator, $\bA$ obeys the quantum Langevin equation,
where the effect of the dissipative
electrons is simulated by a random force.
We will show in what follows that the crucial properties of
$\bA(\br,t)$ used to derive Eq.(\ref{Ktemp})
are valid also for the random vector
potential operator, and hence
  Eq.(\ref{Ktemp}) is valid  when
dissipative bodies are present.
Note, that in case of dissipation normal modes of the EM field do not exist,
the EM field can not be quantized in the usual way,
and this is why one is forced to use the Langevin equation approach.

\section{Quantum Langevin equation for the EM field}
\label{Langeq}
A quantum Langevin equation for the coordinate operator $q$ of a particle
moving in potential $V(q)$, derived in Ref. \cite{FLC88},
can be written in terms of the
particle Lagrangian $L=m\dot{q}^2/2+V(q)$ as
\be
\label{Leqpar}
\frac{d}{dt}\frac{\partial L}{\partial \dot{q}(t)}-
\frac{\partial L}{\partial q(t)}
+\int _{-\infty}^{t} dt'\gamma (t-t')\dot{q}(t')=F(t).
\ee
The kernel $\gamma$ is responsible for the "friction" produced by the
environment,
which is a thermal bath at temperature $T$, and the operator $F(t)$
is the random
force. The statistical and commutation properties of the random force
are defined
by the dissipation kernel  $\gamma$. Namely,
$F(t)$ is a Gaussian stationary random process with $\langle F\rangle=0$
and a correlator
\be
\label{corrF}
\hspace{-2cm}
\frac{1}{2}\left\langle F(t)F(t')+F(t')F(t)\right\rangle
=\frac{1}{2\pi}
\int_{-\infty}^{\infty} d\omega \;\exp[-i\omega (t-t')]
\hbar\omega
\coth\frac{\hbar\omega}{2T}\,\re\gamma(\omega),
\ee
while the commutator of the random force is an imaginary $c$-number,
\be
\label{commF}
[F(t), F(t')]
=\frac{2}{\pi}
\int_{-\infty}^{\infty}  d\omega \exp[-i\omega(t-t')]\;
\hbar\omega\,  \re\gamma(\omega).
\ee
(Note, that this Langevin equation is equivalent to the
well-known approaches used
by Feynman and Vernon \cite{FV63} and Caldeira and Legget \cite{CL83}).

Consider the classical Maxwell equations  for the long-wave EM field,
$
\rot\bE=-\dot{\bH}/c
$ and
$
\rot\bH=\dot{\bD}/c+(4\pi/c){\bf j}
$,
where $\bD$ is the displacement  given by
\be
\bD(\br,t)=\bE(\br,t)+\int_{-\infty}^{t}dt'\;\chi(\br,t-t')\bE(\br,t'),
\ee
and ${\bf j}$ is the external current density.
Note that fields entering the above macroscopic equations
are averaged over a volume $\Delta V=(\Delta L)^3$, where $\Delta L$
is large compared to all relevant microscopic lengths, but small compared
to the wave-length of the EM field.
With the gauge
$\bE=-\dot{\bA}/c$ and $\bH=\rot \bA$
the first Maxwell equation is satisfied
and the second gives an equation for the vector potential

\be
\label{eqA1}
\frac{1}{c^2}\ddot{\bA}(t)&+&\rot\rot \bA(t)+\frac{1}{c^2}
\int dt' \dot{\bA}(t')\dot{\chi}(t-t')
=\frac{4\pi}{c}\bj(t).
\ee
Starting from the EM field Lagrangian one can prove that
 this equation
can be considered as the quantum Langevin  equation for the  random field
operator $\bA$,
 if $\bj$ is the appropriate random force created by the thermal bath
of dissipative electrons.
($j_{\alpha}\Delta V/c$ plays the role of $F$
and $\dot{\chi}\Delta V/4\pi c^2$ plays the role of $\gamma$.)
The correlator of this force is known from the fluctuation-dissipation
theorem \cite{LL},
\be
\hspace{-1.7cm}
\frac{1}{2}\left\langle j_{\alpha}(\br,t)j_{\beta}(\br',t')+
j_{\beta}(\br',t')j_{\alpha}(\br,t)\right\rangle
=\int d\omega \;e^{-i\omega (t-t')} (j_{\alpha}(\br)j_{\beta}(\br'))_{\omega}
\ee
with
\be
\label{jjfr}
(j_{\alpha}(\br)j_{\beta}(\br'))_{\omega}
=\delta_{\alpha\beta}\delta(\br-\br')
\frac{\hbar}{8\pi^2}\;
\omega^2\coth\frac{\hbar\omega}{2T}\;\im\epsilon(\br,\omega).
\ee
Comparing this correlator with Eq.(\ref{corrF}), we can find from
Eq.(\ref{commF})
the commutator of the random currents, which turns out to be an imaginary
$c$-number,
\be
\hspace{-1.2cm}
[j_{\alpha}(\br,t),j_{\beta}(\br',t')]
=\delta_{\alpha\beta}\delta(\br-\br')
\frac{\hbar}{2\pi^2}\int d\omega e^{-i\omega(t-t')}\omega^2
\;\im\epsilon(\br,\omega).
\ee
(The quantum Langevin equation for the EM field
was considered also in Ref.\cite{GW96},
but in a form not suitable for our problem.)

The retarded Green function, corresponding to Eq.(\ref{eqA1}), obeys
\be
\label{Gt}
\hspace{-1.7cm}
\frac{1}{c^2}
\ddot{\cG}_{\alpha\lambda}(\br,\br'; t)+
\rot\rot_{\alpha\beta}\cG_{\beta\lambda}(\br,\br'; t)
+\frac{1}{c^2}
\int dt' \dot{\cG}_{\alpha\lambda}(\br,\br'; t')\dot{\chi}(t-t')
\\ \nonumber
=-4\pi\delta_{\alpha\lambda}\delta(\br-\br')\delta(t),
\ee
with the condition $\cG(t)=0$ for $t<0$.
(The $\rot$ operators are defined
in terms of the antisymmetric tensor $\delta_{\alpha\sigma\beta}$ as
$\rot
 _{\alpha\beta}=
\delta_{\alpha\sigma\beta}\nabla_{\sigma}$ and $
\rot\rot_{\alpha\beta}=
\nabla_{\alpha}\nabla_{\beta}-\delta_{\alpha\beta}\nabla^2$.)
Using this Green function
one can calculate the random field operator
\be
\label{Ajt}
A_{\alpha}(\br,t)=
-\frac{1}{c}\int dt'\int d\br'\cG_{\alpha\beta}(\br,\br';t-t')
j_{\beta}(\br',t').
\ee
Two important consequences follows from this relation.
First, as the commutator of the currents is an imaginary $c$-number
and the $\cG$ is real, the commutator of the random
field operators is also an imaginary $c$-number.
Second, as the current is a stationary Gaussian process
and $\cG$ depends on $t-t'$,
same is the random field operator. Since these two properties of the
field operator
were crucial for deriving the dephasing factor as given by Eq.(\ref{Ktemp})
for the case of no dissipation, we proved thereby
that this result is valid also when dissipation is present.

The last equation allows also the calculation of the correlator
and the commutator of the field operator. The correlator
is known \cite{AGD} to be related to the Green function defined by Eq.(\ref{Gt}),
\be
\label{DG}
\cD_{\alpha\beta}(\br,t;\br',t')
=\int d\omega \;e^{-i\omega (t-t')}
(A_{\alpha}(\br)A_{\beta}(\br'))_{\omega}
\ee
with
\be
\label{Dfr}
(A_{\alpha}(\br)A_{\beta}(\br'))_{\omega}=-\frac{\hbar}{\pi}
\coth\frac{\hbar\omega}{2T}\;
\im \cG_{\alpha\beta}(\omega;\br,\br'),
\ee
where the Green function in the frequency domain is defined as
\be
\cG_{\beta\lambda}(\br,\br'; t)&=&\int
\frac{d\omega}{2\pi}e^{-i\omega t}
\cG_{\beta\lambda}(\omega;\br,\br').
\ee
The definition of $\cG(\omega)$ corresponds to
that in \cite{MM75}.
The Green function is symmetric,
$\cG_{\beta\lambda}(\omega;\br,\br')
=\cG_{\lambda\beta}(\omega;\br',\br),$
and it follows from the properties of $\epsilon(\br,\omega)$ that
$\cG_{\beta\lambda}(\omega;\br,\br')^*=
\cG_{\beta\lambda}(-\omega;\br,\br').$
One can prove the following important integral relation
\be
\label{GG=G}
\hspace{-1cm}
\frac{\omega^2}{c^2}\int d\br_{0}\im \epsilon(\br_{0},\omega)
\cG_{\lambda\alpha}(\omega;\br,\br_{0})
\cG_{\lambda'\alpha}(\omega;\br',\br_{0})^*
=
-4\pi\im \cG_{\lambda\lambda'}(\omega;\br,\br'),
\ee
which  is an obvious generalization to 3D of the
1D relation given in \cite{GW96}.
(It can be proven using the following Green theorem:
$
\int d\br\;\phi \;\rot\rot_{\alpha\beta}\psi=
\int d\br\;\rot_{\sigma\alpha}\phi\;\rot_{\sigma\beta}\psi.
$). Using this integral relation one can obtain Eq.(\ref{Dfr})
and also calculate the commutator
\be
\label{commA}
[ A_{\alpha}(\br)_{\omega},A_{\beta}(\br')_{\omega'}]=-\delta(\omega+\omega')
\frac{2\hbar}{\pi^2}\im\cG_{\alpha\beta}(\omega;\br,\br').
\ee

It is important to notice that deriving Eq.(\ref{Dfr}) and Eq.(\ref{commA})
with the help of Eq.(\ref{GG=G}) one has to
assume  that the temperature (entering Eq.(\ref{jjfr}))
is constant over the whole space (where $\im\,\epsilon \neq 0$).
 This is correct in thermal equilibrium, when all absorbing bodies
 are at the same temperature, and hence
 Eq.(\ref{Dfr}) provides the correlator for the {\it equilibrium EM field}.
But it does not provide the correlator for the
{\it nonequilibrium thermal EM radiation}, when there is radiation energy
exchange between bodies with different temperatures.
This correlator can be also calculated using Eq.
(\ref{jjfr}), but the integral over the source point can not
be simplified using Eq.(\ref{GG=G}).
\section{Dephasing by an infinite  thick plate}
\label{thickpl}
In what follows we consider a simple geometry when the dephasing body
is a half-space $z<0$ with $\epsilon(\br,\omega)=\epsilon(\omega)$
and the two beams move in vacuum $z>0$ in a plane $z=d$ parallel to
the interface.
In this geometry the Green function of the EM field is
translational invariant in the $x,y$  plane and hence can be presented
as follows
\cite{MM75}
\be
\label{Gfe}
\cG_{\alpha\beta}(\omega;\br,\br')=
\int\frac{d^2 k}{(2\pi)^2}\;e^{i\bk(\bR-\bR')}g_{\alpha\beta}(\omega,\bk|z,z'),
\ee
where $\bR$ is the component of $\br$ in the $(x,y)$ plane
and $\bk$ is a vector in this plane.
Because of the special geometry  we are interested in the
Green function $\cG$ for $z=z'=d$ and $\alpha,\beta=x,y$,  and will denote
$g_{\alpha\beta}(\omega,\bk|d,d)\equiv g_{\alpha\beta}(\omega,\bk)$.
Using the explicit expressions for $g_{\alpha\beta}$ given in \cite{MM75},
one can write
\be
\label{g=t+l}
g_{\alpha\beta}(\omega,\bk)=
g_{t}(\omega,k)
\left[\frac{k_{\alpha}k_{\beta}}{k^2}-\frac{1}{2}\delta_{\alpha\beta}\right]+
\frac{1}{2}\delta_{\alpha\beta}\;g_{l}(\omega,k),
\ee
where
\be
\label{g/symm}
\hspace{-1.7cm}
g_{l,t}(-\omega,k)=g_{l,t}(\omega,k)^*,\qquad
g_{\alpha\beta}(\omega,-\bk)=g_{\alpha\beta}(\omega,\bk)=
g_{\alpha\beta}(-\omega,\bk)^*.
\ee
The $g_{\alpha\beta}(\omega,\bk)$
are related to the correlator of the tangential
components of the electric field in the plane $z=d$.
Using Eqs.(\ref{g/symm}) one can check from Eq.(\ref{Dfr})
that for $\alpha,\beta=x,y$
\be
\label{Etcorr/R}
(E_{\alpha}(\bR)E_{\beta}(\bR'))_{\omega}=\int d^2 k\;e^{i\bk(\bR-\bR')}
(E_{\alpha}E_{\beta})_{\omega\bk}
\ee
with
\be
\label{Etcorr/k}
(E_{\alpha}E_{\beta})_{\omega\bk}=
\frac{2\hbar}{(2\pi)^3}\left(\frac{\omega}{c}\right)^2
\coth\frac{\hbar\omega}{2T}\;(-\im g_{\alpha\beta}(\omega,\bk)).
\ee

The classical current in beam 1 is
$
j_{1}(\br,t)=e\bv_{1}(t)\delta(\bR-\bR_{1}(t))\delta(z-d),
$
where $\bR_{1}(t)$ is the trajectory of beam 1 and
$\bv_{1}(t)=d\bR_{1}(t)/dt$ is the
electron velocity in this beam. One can present
\be
\label{cfe}
\bj_{1}(\br,t)
=e\delta(z-d)\int\frac{d^{2}k}{(2\pi)^2}\int d\omega\;
e^{-i\omega t+i\bk\bR}\;(\bl_{1})_{\bk\omega},
\ee
where the radiation amplitude is
\be
\label{vfc}
(\bl_{1})_{\bk\omega}=\int\frac{dt}{2\pi}
\bv_{1}(t)\;e^{i\omega t-i\bk\bR_{1}(t)}.
\ee
The relevant frequencies and wave vectors
are those of the EM field created by the electron in beam 1.
Similar expressions can be written for beam 2.
We now introduce into the dephasing integral, Eq.(\ref{Ktemp}),
the correlator $\cD$
expressed in terms of $\cG$ according to Eq.(\ref{DG}),
and the Fourier expansions of $\bj _{1,2}$ and $\cG$ according to
Eq.(\ref{cfe}) and Eq.(\ref{Gfe}), we find, using Eqs.(\ref{g/symm}),
\be
\label{Kp}
\hspace{-1.6cm}
K=\frac{e^2}{2\hbar c^2}
\int_{0}^{\infty}\frac{d\omega}{2\pi}
\coth\frac{\hbar\omega}{2T}
\int\frac{d^{2}k}{(2\pi)^2}
\left\{-\im  g_{\alpha\beta}(\omega,\bk)\right\}
\;[(l_{\alpha})^{*}_{\bk \omega}(l_{\beta})_{\bk \omega}+c.c.],
\ee
where $(\bl)_{\bk\omega}=(\bl_{1})_{\bk\omega}-(\bl_{2})_{\bk\omega}$.
The contribution to this integral comes from frequencies and
wave vectors which are present simultaneously in the fluctuation
spectra $g(\omega,\bk$) and the radiation spectra $(\bl)_{\bk\omega}$.
Using Eq.(\ref{g=t+l}) one can rewrite the integral over $d^2k$
 as follows
\be
\label{int/k}
\hspace{-1cm}
\frac{1}{2\pi}
\int_{0}^{\infty}dk k
\left\{
-2\im g_{t}(\omega,k)\langle|\hat{l}_{\bk\omega}|^2\rangle
 -
\im[ g_{l}(\omega,k)- g_{t}(\omega,k)]\langle|(\bl)_{\bk\omega}|^2\rangle
\right\},
\ee
where $\langle ...\rangle$ means angular average and
$\hat{l}_{\bk\omega}=\bk(\bl)_{\bk\omega}/k$.

For slow enough electrons one can use the dipole approximation (DA), when
 the term $\bk\bR_{1}(t)$ in Eq.(\ref{vfc}) can be neglected.
The condition for this is $\omega\gg kv$, where $\omega$ and $k$ are
the typical frequency and wave vector of the EM fluctuations
contributing to the integral $K$, and $v$
is the characteristic electron velocity. In this approximation
$(\bl_{1})_{\bk\omega}\equiv (\bl_{1})_{\omega} $,
and $e(\bl_{1})_{\omega}$ is the radiating dipole moment. Now
$(\bl)_{\bk\omega}=(\bl_{1})_{\omega}-(\bl_{2})_{\omega}
\equiv(\bl)_{\omega}$.
We substitute $(\bl)_{\omega}$ into Eq.(\ref{Kp}) and as a result
in the DA the dephasing integral is
\be
\label{Knrel}
K=\frac{e^2}{2\hbar c^2}\int_{0}^{\infty}
\frac{d\omega}{2\pi}
\coth\frac{\hbar\omega}{2T}\;S(\omega)\;|(\bl)_{\omega}|^2,
\ee
where
\be
|(\bl)_{\omega}|^2=
|(l_{x})_{\omega}|^2+|(l_{y})_{\omega}|^2,\qquad
S(\omega)= -\im \int\frac{d^2 k}{(2\pi)^2}g_{l}(\omega,k).
\ee
$S(\omega)$ is related to average amplitude of the
tangential electric field at distance $d$ from the interface.
From Eq.(\ref{Etcorr/R}) at $\bR'=\bR$ one finds
\be
(E^2_{t})_{\omega}=
(E^2_{x}+E^2_{y})_{\omega}=\frac{\hbar}{\pi}\left(\frac{\omega}{c}\right)^2
\coth\frac{\hbar\omega}{2T}\;S(\omega).
\ee
It is convenient to represent $K=K_{p}+K_{e}$, where the two terms
are the contributions to the integral of the domains, correspondingly,
$k<\omega/c$ and $k>\omega/c$.    In the first domain
the wave vector component perpendicular to the interface
$k_{z}=\sqrt{(\omega/c)^2-k^2}$ is real, which
means that this domain corresponds to waves
propagating perpendicular to the interface (PW), while in the second one
$k_{z}$ is imaginary and it corresponds to evanescent waves (EW).

\section{Spectral densities \lowercase{$ g(\omega,k)$}}
\label{SpDensg}
Using the explicit expressions for $g_{\alpha\beta}$ given in \cite{MM75},
 one can find (for $\omega>0$)
\be
\label{glt}
g_{l,t}(\omega,k)=
\frac{2\pi i}{k_{0}}[e^{ipu(\xi)}F_{l,t}(\xi)-G_{l,t}(\xi)],
\ee
where $\xi=|\bk|/k_{0}$ with $k_{0}=\omega/c$,  and
\be
G_{l,t}(\xi)=u\pm\frac{1}{u},\qquad
 F_{l,t}(\xi)=
\pm \frac{1}{u}\;\frac{v-u}{v+u}-u\;
\frac{v-\epsilon(\omega)u}{v+\epsilon(\omega)u},
 \ee
 with $
u=[1-\xi^2]^{1/2},\; v=[\epsilon(\omega)-\xi^2]^{1/2},\im u>0,\;\im v >0$
and $p=2k_{0}d$.
The upper sign corresponds to $l$ and the lower to $t$.

\begin{figure}
\vspace{-1cm}
\centerline{\psfig{figure=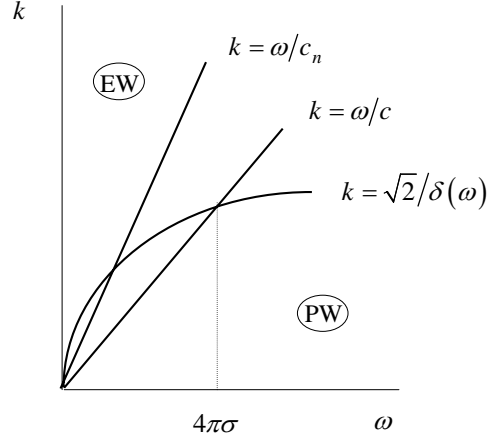,width=10cm}}
\vspace{-6cm}
\caption{Borderlines in the $(\omega, k)$ plane, see text.}
\label{fig1}
\end{figure}

Consider the spectral densities $(-\im g_{l,t})$ in the $(\omega,k)$ plane
at $\omega>0, k>0$ (see Figure \ref{fig1}).
One important borderline in this plane is $\xi=1$, i.e. $k=\omega/c$,
which, as noted above, separates the propagating waves
(PW)  domain below it
from the evanescent waves (EW)  domain above it.
In the PW domain  $\xi<1$ and
\be
-\im g_{l,t}(\omega,k)=\frac{2\pi}{k_{0}}[G_{l,t}(\xi)-
\re F_{l,t}(\xi)\cos pu(\xi)],
\ee
while in the EW domain $\xi>1$ and
\be
\label{gev}
-\im g_{l,t}(\omega,k)=-\frac{2\pi}{k_{0}}e^{-p\sqrt{\xi^2-1}}\re F_{l,t}(\xi).
\ee
For $\epsilon(\omega)\equiv 1$ one finds $u=v$ and $F(\xi)=0$.
This corresponds to empty space, in which case the spectral densities
$-\im g_{l,t}(\omega,k)\neq 0$ only in the PW domain, where in this case
\be
\label{gvac}
-\im g_{l,t}(\omega,k)=\frac{2\pi}{k_{0}}G_{l,t}(\xi).
\ee
The same result is obviously obtained far from the interface,
when $d\rightarrow\infty$, since one can neglect the oscillating
or decaying term in Eq.(\ref{glt}).

The second important borderline
is $\xi=|\epsilon(\omega)|^{1/2}$.
For a non-dispersive lossless dielectric,
$\im \epsilon=0, \;\epsilon\equiv n^2>1$, this borderline is simply
$k=\omega/c_{n}$,
where $c_{n}\equiv c/n$ is the light velocity in the dielectric.
One easily finds that $-\im g_{l,t}(\omega,k)= 0$ above the
second borderline, for $\xi>n$.
Below it, for $1<\xi<n$, one has
\be
\label{F/diel}
\re F_{l,t}(\xi)=-\frac{2}{\epsilon -1}(\epsilon-\xi^2)^{1/2}
\left[\frac{\epsilon (\xi^2 -1)}{(\epsilon +1)\xi^2-\epsilon}\pm 1\right].
\ee
In the generic case of arbitrary complex
$\epsilon(\omega)$ the spectral densities
$(-\im g_{l,t})$ are non-zero in the whole  $(\omega,k)$ plane.

\noindent

In what follows we will consider two cases: (i) a highly polarizable lossless
dielectric, $\im \epsilon=0, \;\epsilon\gg 1$, and (ii) a "good" conductor,
$\im \epsilon\gg\re \,\epsilon\simeq 1$. In the last case we write
$\epsilon(\omega)=\epsilon_{0}+i(4\pi\sigma/\omega)$, where $\sigma$ is
the conductivity, and assume that $\sigma$ is larger than all
relevant frequencies.
 Both cases correspond to $|\epsilon|\gg 1$ and it is
instructive therefore to investigate the limit $|\epsilon|=\infty$.
In this limit  $F(\xi)=G(\xi)$ and
the spectral densities $-\im g_{l,t}(\omega,k)\neq 0$ only in the
PW domain, where
\be
-\im g_{l,t}(\omega,k)=\frac{2\pi}{k_{0}}[1-\cos pu(\xi)]G_{l,t}(\xi).
\ee
One can see that near the interface with an ideal conductor ($\sigma=\infty$)
or an ideal dielectric  ($\epsilon_{0}=\infty$)  the tangential
electric field fluctuations are suppressed.
In the case of a conductor it is obvious from the boundary condition
$E_{t}=0$. In the case of a dielectric this boundary condition is also
effective, since $E_{t}\neq 0$ would mean, due to the continuity of $E_{t}$,
infinite energy density in the dielectric or infinite displacement current.

Now we turn to the spectral density corrections
which are due to finite $|\epsilon|$.
To investigate the role of these corrections we use the following
expansions. Well below the second borderline, i.e. at
 $\xi\ll |\epsilon|^{1/2}$, one has
\be
\label{Fexpll}
\hspace{-1.9cm}
F_{l}(\xi)=G_{l}(\xi)-4\epsilon^{-1/2}+O(\epsilon^{-1}),\qquad
F_{t}(\xi)=G_{t}(\xi)\left[1-2\epsilon^{-1}\right]+O(\epsilon^{-3/2}).
\ee
Well above the second borderline, i.e. at
$\xi\gg |\epsilon|^{1/2}$, one finds
\be
\label{Fexpgg}
F_{l,t}(\xi)=i\;\frac{\epsilon -1}{\epsilon +1}\;\xi
+i\left(\frac{\epsilon -1}{\epsilon +1}\right)^2\frac{1}{2\xi}+O(\xi^{-3}).
\ee
At $\xi\simeq |\epsilon|^{1/2}$ obviously
$F_{l,t}(\xi)\simeq |\epsilon|^{1/2}$.

It follows from Eq.(\ref{Fexpll}) that in the PW domain
the finite $|\epsilon|$ corrections are small. However as we will see later,
these corrections are important in the EW domain
at small distances from the interface $d$. To account for these corrections
in a dielectric one can use the explicit expressions given
by Eq.(\ref{F/diel}),
but for a conductor the situation is more complicated.
 For a good conductor it is convenient
to use the surface impedance $\zeta(\omega)$ and the skin depth
$\delta(\omega)$  defined as follows,
\be
\hspace{-1.6cm}
\zeta(\omega)=(\omega/8\pi\sigma)^{1/2}
\approx (2|\epsilon(\omega)|)^{-1/2},\qquad
\delta(\omega)=\frac{c}{(2\pi\sigma\omega)^{1/2}}=
2\zeta(\omega)\frac{c}{\omega}\;.
\ee
In these terms the second borderline is $k=\sqrt{2}/\delta(\omega)$
or $\xi=(\sqrt{2}\zeta(\omega))^{-1}$.
Since  $\zeta(\omega)\ll 1$ and $\delta(\omega)\ll k_{0}^{-1}$
when $\omega\ll 4\pi\sigma$,
this borderline in the EW domain is
well above the first  borderline $k=\omega/c$.

In between the borderlines, $\omega/c\ll k\ll \delta(\omega)^{-1}$,
one can find using Eqs.(\ref{Fexpll}),
\be
\re F_{l}(\xi)=-4\zeta(\omega),\qquad \re F_{t}(\xi)=-4\zeta(\omega)^2\xi.
\ee
Above the upper borderline, $k\gg \delta(\omega)^{-1}$, one finds from
Eq.(\ref{Fexpgg}), using
$(\epsilon -1)/(\epsilon +1)=1+4i\zeta(\omega)^2$,
the dominant term to be
\be
\label{F/xi/inf}
\re F_{l,t}(\xi)=-4\zeta(\omega)^2\xi.
\ee
Now we find from Eq.(\ref{gev})  the spectral densities for a good conductor
in the EW domain. In between the borderlines
\be
\label{g/between}
\hspace{-1.7cm}
-\im g_{l}(\omega,k)=\frac{8\pi}{k_{0}}\zeta(\omega)e^{-2kd},
\qquad
-\im g_{t}(\omega,k)=\frac{8\pi}{k_{0}^2}\zeta(\omega)^2 k e^{-2kd},
\ee
while above the upper borderline
\be
\label{g/above}
\hspace{-1.8cm}
-\im g_{l,t}(\omega,k)=\frac{8\pi}{k_{0}^2}\zeta(\omega)^2 k e^{-2kd},
\quad
-\im g_{\alpha\beta}(\omega,k)=\frac{8\pi}{k_{0}^2}
\zeta(\omega)^2 k e^{-2kd}\frac{k_{\alpha}k_{\beta}}{k^2}.
\ee
In the EW domain the fluctuations are small, because of the small
surface impedance, and are strongly suppressed at $k\gtrsim d^{-1}$,
since random fields created by
fluctuations of the random currents with wavelength
shorter than $d$ are averaged at distance $d$.

\section{Spectral density $S(\omega)$}
\label{SpDensS}
The spectral density,
which enters in the DA, can be split into contributions
of the PW and EW domains, $S(\omega)=S_{p}(\omega)+S_{e}(\omega)$, with
\be
\label{S/int xi}
S_{p}(\omega)&=&-\frac{1}{2}k_{0}\re \int_{0}^{1}d\xi
\;\xi[e^{ip\sqrt{1-\xi^2}}
F_{l}(\xi)-G_{l}(\xi)],
\\ \nonumber
S_{e}(\omega)&=&
-\frac{1}{2}k_{0}\int_{1}^{\infty}d\xi\;\xi
e^{-p\sqrt{\xi^2-1}}\re F_{l}(\xi).
\ee
To avoid misunderstanding we note that the spectral density due to a
 half-space $z<0$ with $\epsilon (\omega)$
 calculated in Ref. \cite{RKT87} is not for
the equilibrium EM field, but  is  for the radiation into a zero temperature
half space $z>0$.

In the limit $|\epsilon|=\infty$ only PW contribute
and one finds easily the spectral density
\be
\label{Sim}
S(\omega)=S_{p}(\omega)=\frac{\omega}{c}\;
\left[\frac{2}{3}-\frac{\cos p}{p^2}-
\frac{\sin p}{p}\left(1-\frac{1}{p^2}\right)\right].
\ee
At large distances from the interface,
$d\gg k_{0}^{-1}$, one has $S(\omega)=(2/3)k_{0}$, which correspond to EM
field fluctuations in empty space.
(The factor 2/3 appear because only two tangential components of the
electric field are relevant).
 At small distances, $d\ll k_{0}^{-1}$,
the fluctuations are suppressed, $S(\omega)\sim d^2$.
For an ideal conductor or ideal dielectric the contribution to $S(\omega)$
comes from $\xi\simeq 1$ (i.e. $k\simeq \omega/c$)
independent of the value of the parameter $p=2k_{0}d$.

Now we turn to the finite $|\epsilon|$ corrections.
One can see from Eq.(\ref{Fexpll}) that the corrections to $S_{p}(\omega)$
are small, and one can use for $S_{p}(\omega)$ the result given by
Eq.(\ref{Sim}).
But this is not the case for $S_{e}(\omega)$
when $d$ is small and the decay of the exponent is this integral
is slow. For a dielectric, $S_{e}(\omega)$ can be
calculated using Eq.(\ref{F/diel}) and when $\epsilon=n^2\gg 1$ the result is
\be
\label{Sdiel}
\hspace{-1.7cm}
d\ll\lambda _{n}:\quad S_{e}(\omega)=\frac{2}{3}k_{0}n\; ;\qquad
d\gg\lambda _{n}&:&\quad S_{e}(\omega)=
\frac{1}{2k_{0}n}\,\frac{1}{d^2} ,
\ee
where $\lambda _{n}=c_{n}/\omega=(k_{0}n)^{-1}$
is the wave length in the dielectric. The first of Eqs.(\ref{Sdiel})
means simply that the fluctuations of $E_{t}$  at
$d\ll\lambda _{n}$ in the vacuum are the same as in the dielectric,
 which is consistent with the continuity
of $E_{t}$  at the interface.
Eqs.(\ref{Sdiel}) clearly demonstrate
the importance of finite $|\epsilon|$ corrections at small
distances. Comparing $S_{e}(\omega)$ for $d\gg\lambda _{n}$
 with $S_{p}(\omega)$
from Eq.(\ref{Sim}), one can see,
that the ideal dielectric approach, when $S(\omega)$ is dominated by
PW,  is valid only at $d\gg \lambda n^{-1/4}=\lambda_{n} n^{3/4}$,
while at smaller distances $S(\omega)$ is dominated by EW.
 According to the ideal dielectric approximation, Eq.(\ref{Sim}),
near the interface the fluctuations are suppressed compared
to free space,
while from the first of Eqs.(\ref{Sdiel}) it follows that they are
enhanced compared to free space by a factor $n$.
The contribution to the integral $S_{e}(\omega)$
comes from $k\simeq\lambda_{n}^{-1}$, when $d\ll\lambda_{n}$, and from
$k\simeq d^{-1}$  when $d\gg\lambda_{n}$. When $S_{e}(\omega)$
dominates, both cases correspond
to large imaginary $k_{z}$. In other words, the fluctuations
contributing to $S_{e}(\omega)$,
are due to near fields localized close to the interface.

The situation in a conductor is different, since from Eq.(\ref{F/xi/inf})
one can see that the  integral $S_{e}(\omega)$ diverges at $d=0$.
The near-field fluctuations in the case of a conductor are
\be
\label{Scond}
\hspace{-2.5cm}
d\ll\delta(\omega):\; S_{e}(\omega)=\frac{\zeta(\omega)^2}{2k_{0}^2 d^3}=
\frac{\delta(\omega)^2}{8 d^3};
\quad
d\gg\delta(\omega):\; S_{e}(\omega)=\frac{\zeta(\omega)}{2k_{0} d^2}=
\frac{\delta(\omega)}{4 d^2}.
\ee
The near field/far field crossover
point is $d_{\times}\simeq(c/\omega)\zeta(\omega)^{1/4}$ and for all $d$
 the relevant $k\simeq d^{-1}$.
When $d\ll\delta(\omega)$ the relevant $k$  are above the
borderline  $k=\sqrt{2}/\delta(\omega)$
and the first of Eqs.(\ref{Scond}) is obtained using Eq.(\ref{g/above}),
while when $d\gg\delta(\omega)$ the relevant $k$ are below this
borderline and the second of Eqs.(\ref{Scond}) is obtained using
Eq.(\ref{g/between}).

Comparing $S_{e}(\omega)$ for a dielectric and a conductor one can see
that they are similar not very close to the interface,
if one replaces $n^{-1}$ by $\zeta(\omega)$ and $\lambda_{n}$ by
$\delta(\omega)$. The behavior very close to the interface is different,
since for a dielectric $S_{e}(\omega)$ is finite, while for a conductor
it diverges.
(This singularity is cut-off if one takes into account the spatial dispersion
of the conductor $\epsilon$ \cite{RKT87}).

\section{Dephasing by near fields}
\label{Deph/nf}
\begin{figure}
\vspace{-1cm}
\centerline{\psfig{figure=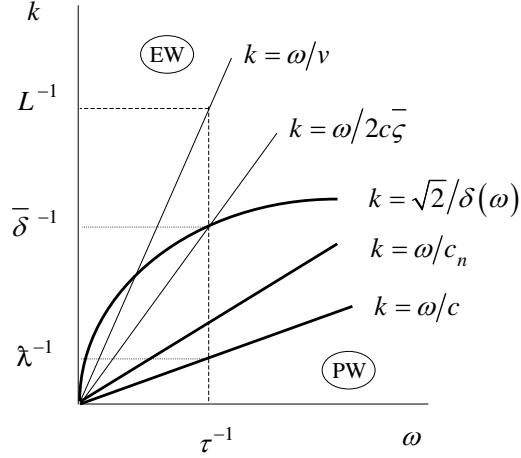,width=9cm}}
\vspace{-4.5cm}
\caption{Integration domains in the $(\omega, k)$ plane, see text.}
\label{fig2}
\end{figure}

To simplify the picture of dephasing we  assume that
the electrons  smoothly accelerate and decelerate. In other words,
we assume that
the electron motion has only one characteristic time scale $\tau$,
which is the time of flight from the emitter to the detector, and
only one length scale $L$, which is the trajectory length.
The characteristic electron velocity is defined as $v=L/\tau$.
We also assume, following the experimental situation, that the beams
are close, i.e. the distance between them $a$ is small compared to $L$.
The  frequencies of the EM waves emitted by the electrons are of the
order of $\omega=\tau^{-1}$ and wavelengths are of the order of
$\lambda=c\tau$.
 In this  model the angle
averages entering the integrals in Eq.(\ref{int/k})
and Eq.(\ref{Knrel}) can be presented as follows:
\be
\label{Psi}
\hspace{-2cm}
\langle|\hat{l}_{\bk\omega}|^2\rangle=\theta L^2\Psi_{1}(z,y),\quad
\langle|(\bl)_{\bk\omega}|^2\rangle=\theta L^2\Psi_{2}(z,y),\quad
|(\bl)_{\omega}|^2=\theta L^2\Psi_{2}(z,0),
\ee
where $z=\omega\tau/2$, $y=kL$, and the small factor $\theta=(a/L)^2$
appears because the beams are close and the effective current $\bj _{12}$
is smaller than the beam currents $\bj _{1}$ and $\bj _{2}$.
The functions $\Psi$ decay fast enough at $\omega\gtrsim\tau^{-1}$
and $k\gtrsim L^{-1}$, restricting the integration in the $(\omega,k)$
plane in Eq.(\ref{Kp}) within the rectangle
$\square\equiv[\;0<\omega\lesssim\tau^{-1},\; 0<k\lesssim L^{-1}]$,
shown in Figure \ref{fig2} by dashed lines.

The frequency integral in Eq.(\ref{Knrel})
 contains three characteristic frequencies, namely,
$T/\hbar$ (the frequency of the EM field thermal fluctuations),
$\tau^{-1}$ (the frequency radiated by the electron),
 and $c/d$ (the frequency which enters the spectral densities $g$).
Assuming $v/c=0.1$ and $L=10$ cm, we have $\tau=3\times 10^{-9}$ s.
The electro-optical system, which creates, guides and detects the beams,
is at room temperature, and this is the temperature of the EM field
surrounding the beams.
At room temperature $\hbar/T=2.5\times 10^{-14}$ s and obviously
always $T/\hbar\gg \tau^{-1}$. Hence
one can replace in the integral Eq.(\ref{Kp})
$\coth(\hbar\omega/2T)$
by its classical high temperature approximation.

From the spectral densities calculated in sec.\ref{SpDensg} and
sec.\ref{SpDensS} it follows that when $|\epsilon|\gg 1$ the
contribution $K_{p}$ of the PW can be calculated as for an ideal conductor
or dielectric. To calculate this contribution one can employ
the DA, Eq.(\ref{Knrel}), because when  $|\epsilon|=\infty$
the density $S(\omega)$ comes from $k\simeq \omega/c$, and
for non-relativistic electrons
the condition for the DA to be valid, namely $kv\ll \omega$,
is satisfied.

First we calculate the
dephasing near an ideal conductor or ideal dielectric, given by $K_{p}$,
 Eq.(\ref{Knrel}). Substituting there
 $|(\bl)_{\omega}|^2$ from Eq.(\ref{Psi})
we obtain
\be
\label{K/ffonesc}
d\gg\lambda&:&\quad
K_{p}=K_{0}=
b_{p1}\,\alpha\theta\left(\frac{L}{c}\right)^2\frac{T/\hbar}{\tau}\; ;
\\ \nonumber
d\ll\lambda&:&\quad
K_{p}=b_{p2}\,K_{0}\frac{d^2}{\lambda^2}.
\ee
Here $K_{0}$ is the dephasing in free space, $\alpha=e^2/\hbar c$ is
the fine structure constant and
 the numerical factors $b\simeq 1$ are given in terms of
integrals
\be
b_{p1}=\frac{2J_{0}}{3\pi},\; b_{p2}=\frac{16J_{2}}{5J_{0}};\quad
J_{k}=\int_{0}^{\infty}dz z^k\Psi_{2}(z,0).
\ee
The second of Eqs.(\ref{K/ffonesc}) demonstrates that near an ideal conductor
and dielectric the dephasing is suppressed.
For the parameters used one obtains, neglecting numerical factors,
$K_{0}\simeq 10\theta$. This means that for well separated beams
$(\theta\simeq 1)$ at room temperature the dephasing due to
thermal fluctuations can be significant. In the existing experiments,
however, the beams are very close, $a=100\mu$m, and the
dephasing is negligible, $K_{0}\simeq 10^{-5}$.

Now we turn to EW contribution $K_{e}$, which is responsible for
the enhancement of the dephasing near the interface.
As one can see from Figure \ref{fig2},
 this  contribution is large for small electron
velocities $v$, when the overlap of the rectangle $\square$ with
the EW domain is maximal. Motivated by this we consider
first the simpler case of dielectric for $v\ll c_{n}=c/n$.
Since for a dielectric the spectral densities
$-\im\,g_{l,t}(\omega,k)$ vanish above the borderline
$k=\omega/c_{n}$,
 the integration domain overlaps only with the "bottom" of the rectangle
$\square$, where $k$ is small,
 meaning that one can employ the DA.
Being interested in an almost ideal dielectric ($n\gg 1$), we
substitute the spectral density from Eqs.(\ref{Sdiel}) into
Eq.(\ref{Knrel}) and find the dephasing due to near fields to be
\be
\label{Knfdiel}
 d\ll \lambda_{n}&:&\quad K_{e}=\frac{1}{2}K_{0}n;\\ \nonumber
d\gg \lambda_{n}&:&\quad K_{e}=b_{e}\,K_{0} \frac{\lambda
^2}{n\,d^2}, \quad b_{e}=\frac{3 J_{-2}}{8 J_{0}},
\ee
where $\lambda_{n}\equiv c\tau/n$.
Comparing the second of Eqs.(\ref{Knfdiel}) with
Eq.(\ref{K/ffonesc}) one finds the far field - near field
crossover for dephasing to be $d_{\times}\simeq\lambda n^{-1/4}$,
obviously the same as for $S(\omega)$.
 Contrary to the predictions of the ideal
dielectric approximation $(n=\infty)$ the dephasing of beams
moving near the interface is not suppressed compared to empty
space, but enhanced by a
large factor $n/2$.
Since for most dielectrics $n$ does not exceed 10, the condition
$v\ll c/n$ is not very severe for non-relativistic electrons, but
on the other hand, the enhancement of the dephasing near the interface
is not very strong.

The situation is much more complicated for conductors, since the
spectral densities $-\im\,g_{l,t}(\omega,k)$ do not vanish above
the upper borderline $k=\sqrt{2}/\delta(\omega)$ and the
integration domain overlaps with the whole rectangle $\square$.
 The parameter which plays the role of $1/n$ in case
of a conductor is $\zb=(8\pi\sigma\tau)^{-1/2}$, i.e. the surface
impedance calculated for the characteristic frequency $\tau^{-1}$.
Copper at room temperature has
$\sigma=5\times 10^{17}$sec$^{-1}=6\times 10^{5}\,(\Omega {\rm cm})^{-1}$,
so for a good conductor this
 impedance can be as small as $10^{-5}$,
and hence the restriction $v\ll\zb c$ can be
very severe. As a result one has to consider velocities
larger than $\zb c$, when the DA might be invalid.
Consequently the calculations are
very involved, so we first present the results, discuss them,
and sketch the calculations in the Appendix. In what follows we present
the results for $\kappa\equiv K/K_{0}$ and
the crossover distance $d_{\times}$ from near-field to far-field dephasing.
The results are given in terms of the trajectory length $L$, the radiated
wave length $\lambda=c\tau$, the surface
impedance $\zb$, and the skin depth $\db=2\lambda\zb$.
All numerical factors of order one are omitted.

Three velocity intervals are relevant, namely,
\be
\label{vint}
\hspace{-1.7cm}
A:\quad v/c\ll\zb;\qquad
B:\quad \zb\ll v/c\ll \zb^{1/4};\qquad
C:\quad\zb^{1/4}\ll v/c.
\ee
The crossover distances from near-field to far-field dephasing
in these intervals are as follows
\be
\label{dcross}
A+B:\quad d_{\times}= \zb^{1/4}\lambda; \qquad C:\quad  d_{\times}=
\zb^{1/2}(\lambda^2/L).
\ee
The far-field dephasing, at $d\gg d_{\times}$, in all velocity intervals
 is given by Eq.(\ref{K/ffonesc}).
The near-field dephasing, at $d\ll d_{\times}$, is different
in different velocity intervals. \newline In interval $A$
\be
\label{A}
\hspace{-2cm}
d\ll L:\; \kappa=\db^2\lambda/L^3;\quad
L\ll d\ll \db :\;\kappa=\db^2\lambda/d^3;\quad
\db\ll d\ll d_{\times}:\;\kappa =\db\lambda/d^2.
\ee
In interval $B$:
\be
\label{B}
d\ll L:\quad \kappa=\db\lambda/L^2;\qquad
L\ll d\ll d_{\times} :\quad\kappa=\db\lambda/d^2.
\ee
In interval $C$:
\be
\label{C}
 d\ll d_{\times} :\quad \kappa=\db\lambda/L^2.
\ee
As one can see from the above results,
in the velocity interval $A+B$ the crossover $d_{\times}$ is the same
as for $S(\omega)$ and $L\ll d_{\times}\ll \lambda$.
In this velocity interval $K$ depends on
$d$ in a non-monotonous way, reaching a minimum at $d_{\times}$, where
$\kappa\simeq \zb\,^{1/2}$.
In the velocity interval $C$ one finds $\db\ll d_{\times}\ll L$
and approaching the interface
$K$ decays monotonously till $d\simeq d_{\times}$, where it saturates
at $\kappa\simeq \zb (v/c)^{-2}$.
For all velocities $K$ is finite at $d=0$, since at very small $d$ the
DA is invalid, and the singularity  $d^{-3}$
in $S(\omega)$ is cut-off by the ineffectiveness of wave vectors
$k\gtrsim L^{-1}$.

As was already mentioned, the dephasing $K_{0}$ in empty space is very weak,
and this is why the possible enhancement of $K$ near the interface
is of special interest.
Looking for the ratio $\eta\equiv K(d=0)/K_{0}$ one can see from the above
results
that the dephasing near the interface
is enhanced compared to that in empty space
 only for small enough electron velocities, when $v/c\ll \zb\,^{1/2}$, i.e.
in the interval  $A$ and in the smaller velocity part of interval $B$.
Since in the experiment the fixed parameter is not $\tau$, but $L$,
and hence $\zb$ depends on $v$, it is more convenient to use a
different parameter, namely $\gamma=c/8\pi\sigma L$. In terms of this parameter
the velocity interval $A$ is $v/c\ll\gamma$ and the dephasing enhancement
in this interval is $\eta=\gamma(v/c)^{-2}$.
The smaller velocity part of interval $B$ is
$\gamma\ll v/c\ll\gamma\,^{1/3}$ and here
$\eta=\gamma\,^{1/2}(v/c)^{-3/2}$.
(Note also, that the necessary condition $\zb\ll 1$ reduces to
$\gamma (v/c)\ll 1$ and is always satisfied when $\gamma\ll 1$.)
For the parameters used above one finds $\gamma\simeq 10^{-10}$ and
$\gamma\,^{1/3}\simeq 10^{-3}$. It is clear now that
in the case of a good metal the dephasing
is enhanced only for relatively slow electrons and is not very high.
For example, when $v/c=10^{-4}$ one finds $\eta\simeq 10$.
Much stronger dephasing can be achieved with a high resistivity
semiconductor, for example Si with
$\sigma=1\,(\Omega {\rm cm})^{-1}$, in which case $\gamma\simeq 10^{-4}$
and for $v/c=10^{-4}$ one finds $\eta\simeq 10^{4}$.

\section{Dissipation versus dephasing}
\label{diss-deph}
The coherence of the electrons in the beams can be destroyed
only if there are mechanisms which allow their energy to be dissipated.
When there are no absorbing bodies in the EM environment of the beams,
Eq.(\ref{Ktemp}) describes dephasing related to the dissipation
of electron energy by radiation of EM waves "to infinity".
In fact it means that the energy is dissipated in very far bodies,
not included in the consideration explicitly.
Eq.(\ref{Ktemp}) is formally valid also when the beams are within a
lossless cavity,  if the small
absorbtion in the walls is still large enough to prevent EM field buildup
in the cavity.

If electrons in the two beams move along close trajectories and with
similar velocities, $\bj_{12}=\bj_{1}-\bj_{2}$ is small and the dephasing is weak.
When the distance between the trajectories is small compared
to the correlation length of the EM field in the direction
perpendicular to the beams,
the random electric fields in adjacent points of the two
trajectories  fluctuate synchronously, and as
a result electrons in both beams change their phases also
synchronously, which means that the beams remain mutually coherent.
It does not mean, however, that the energy losses in the beams,
defined by the currents $\bj_{1}$ and $\bj_{2}$ separately,
are small.
There is one additional very important difference between dephasing
and dissipation. Using the relation
$
\langle D(\alpha)^{\dag} a^{\dag} a D(\alpha)\rangle_{T}=n+|\alpha|^2
$, where $D(\alpha)=\exp[\alpha a^{\dag}-\alpha^{*}a]$, one can prove
that the energy radiated by a {\it classical} current into a
thermal EM field is
\be
\hspace{-1.7cm}
W=\frac{1}{2\hbar c^2}\int dt dt'\int d\br d\br'
j^{\alpha}(\br,t) j^{\beta}(\br',t')\;
i\frac{\partial}{\partial t}[ A_{\alpha}(\br,t),\;A_{\beta}(\br',t')].
\ee
 Hence, in strong
contrast to dephasing, the energy losses of the beam electrons
do not depend on the environment temperature.

\ack
I acknowledge the discussions with
F. Hasselbach and P. Sonnentag regarding the experimental situation.
I would like to thank  Y.Imry and  A.Stern for discussions
related to a similar dephasing problem in solid state physics
and P.W\"{o}lfle for discussions clarifying the classical
approximation for beam currents.
 This work was supported by the Alexander von Humboldt Foundation
and  by the Center of Excellence of the Israel
Science Foundation, Jerusalem.

\section*{Appendix}
In what follows we sketch the calculations of the results presented
in Eqs.(\ref{dcross}), (\ref{A}),(\ref{B}) and (\ref{C}).
There are two contributions to $K_{e}$,
namely $K_{e}''$, coming from above the borderline
$k=\sqrt{2}/\delta(\omega)$, and $K_{e}'$, coming from between
the borderlines $k=\omega/c$ and $k=\sqrt{2}/\delta(\omega)$.
These two contributions  can be estimated using
$-\im\,g_{l,t}(\omega,k)$ from Eqs.(\ref{g/above}) and
(\ref{g/between}), correspondingly.  One can
also see from Fig.\ref{fig2} that when $d\gg L$ the cut-off factor
$e^{-2kd}$ in $-\im\,g_{l,t}(\omega,k)$ selects from the rectangle
$\square$ only its "bottom" and hence the DA is
valid.

In the velocity interval $A$ the lengths hierarchy is $L\ll \db\ll \lambda$.
When $d\ll\db$ the main contribution is $K_{e}''$,
where $g_{l}=g_{t}$, and the second term in Eq.(\ref{int/k}) vanishes.
As a result
\be
K_{e}=K_{e}''=
\frac{3}{J_{0}}K_{0}\frac{\db^2\lambda}{L^3}
\int_{0}^{\infty}\frac{dz}{z^2}\int_{0}^{\infty}dy y^2
e^{-(2d/L)y}\Psi_{1}(z,y).
\ee
If $d\gg L$ one can put $z=0$, which corresponds to the
DA, in agreement with what was stated above, and obtain
\be
K_{e}=\frac{3J_{-2}}{4J_{0}}K_{0}\frac{\db^2\lambda}{d^3}.
\ee
If $d\ll L$ one can put $d=0$ and the integral is a numerical factor
of order one.
When $d\gg\db$ the main contribution is $K_{e}'$
with $g_{l}$ dominating,
and in addition the DA is valid.
Substituting the second of Eqs.(\ref{Scond}) into Eq.(\ref{Knrel})
one finds
\be
K_{e}=K_{e}'=\frac{3J_{-3/2}}{2^{7/2}J_{0}}K_{0}\frac{\db\lambda}{d^2}.
\ee

In the velocity interval $B+C$
the lengths hierarchy is $\db\ll L\ll \lambda$.
Here the main contribution is always $K_{e}'$
and one can check that $g_{t}$ can be neglected compared to $g_{l}$.
As a result
\be
K_{e}=K_{e}'=\frac{3}{2^{1/2}J_{0}}K_{0}\frac{\db\lambda}{L^2}
\int_{0}^{\infty}\frac{dz}{z^{3/2}}\int_{0}^{\infty}dy y
e^{-(2d/L)y}\Psi_{2}(z,y).
\ee
If $d\gg L$ one can put $y=0$ and obtain
\be
K_{e}=\frac{3J_{-3/2}}{2^{5/2}J_{0}}K_{0}\frac{\db\lambda}{d^2}.
\ee
When $d\ll L$ one can put $d=0$ and the integral is a numerical factor.
These  results are valid in the whole velocity interval $B+C$.
The separation appears when one compare the near-field and far-field
contributions and finds that $d_{\times}\gg L$  in $B$ ,
while  $d_{\times}\ll L$ in $C$.

\end{document}